\documentstyle[11pt]{article}
\begin{document}

\title{Exact low-energy landscape and relaxation phenomena in 
ISING spin glasses}

\author{ T. Klotz and S. Kobe \thanks{e-mail address: KOBE @ 
PTPRS1.PHY.TU-DRESDEN.DE} \\
Institut f\"ur Theoretische Physik, TU Dresden, \\
D-01062 Dresden, Federal Republic of Germany}
\date{}
\maketitle

\begin{abstract}
All ground states and low-lying excitations of a $ \pm $ $I$ 
Ising spin glass model on a cubic $ 4 \times 4 \times 4 $ 
lattice with periodical boundary conditions were calculated 
using a method of combinatorical optimization. The structure of 
states in the phase space is enlightened by a representation by 
means of clusters in the configuration space and their 
connectivity.
The relaxation behaviour of the system can be described by 
random walks in the high-dimensional phase space. Rewriting 
this task as an eigenvalue problem the influence of `entropic 
barriers' in comparison with an unstructured system can be 
studied, leading to a better understanding of the anomalous 
slow dynamic behaviour.
\end{abstract}

\section{Introduction}

It is well established that the complicated structure of the 
phase space is the key to understand the unusual behaviour of 
spin glasses. But up to now, there is little known about the 
details of this structure even in the simple case of an Ising 
spin-glass model.

Because of the complexity of the problem and the huge number of  
$ 2^{N} $ configurations of  the system with $N$ spins exact 
complete enumerations restrict the size of the system which can 
be considered to  $N = 20 \ldots 30$. In principle, the ground 
state is given by the minimization task, in the zero-field 
case,
\begin{equation}
E_0=\min_{s_i=1 \vee -1}\left( - \sum_{i<j}{I_{ij} \, s_i \, 
s_j} \right) \, .
\end{equation}
For analyzing the exact low-energy landscape the knowledge of 
postoptimal solutions of (1) is necessary, too. Optimization 
tasks of this kind are also important in other fields of 
physical, mathematical and biological sciences, see e.g. the 
Coulomb glass problem \cite{1,2}, the Travelling Salesman 
problem \cite{3} and the protein folding problem \cite{4}.

The first numerical method based on nonlinear combinatorical 
optimization used a branch-and-bound algorithm \cite{5}. It was 
applied to find the ground state of a two-dimensional amorphous 
Ising system with $N = 40 \ldots 60$ antiferromagnetically 
interacting spins (\cite{6} - \cite{8}). For a $ \pm $ $I$ spin 
glass on a square $ L \times  L $ lattice the ground state can 
be found with polynomial in $N$ increasing computing time 
(\cite{9} - \cite{11}). The first exact analysis of the 
morphology of ground states of a spin system with periodical 
boundary conditions was given by Barahona et al. \cite{10} for 
$L = 20$ using the minimal matching of frustrated plaquettes 
\cite{12}. Systems with equally distributed + $I$ and - $I$ 
interactions are characterized by a considerable ground-state 
degeneration, but there are spins, which maintain the same 
orientation in all ground states and thus form connected 
patches, the so-called `packets of solidary spins' \cite{10}. 
Such a `packet' can be flipped as a whole without any cost in 
energy. An exact method to compute the complete partition 
function for $L$ up to 36 is described in \cite{13}.

For $ \pm $ $I$ systems with increasing concentration of 
antiferromagnetic bonds $p$ the ground-state threshold is 
determined as the critical concentration $ p_{c} $ for the 
vanishing of the ferromagnetic ground state. This concentration 
can be calculated by heuristic algorithms (see e.g. \cite{14, 
15}) or by exact minimal matchings up to a size of $L = 300$ 
\cite{16}.

Two-dimensional problems with external field \cite{17} and 
three-dimensional ones \cite{18} belong to the class of NP-hard 
problems \cite{19}. This makes it unlikely that an algorithm 
can be found which is as efficient as in the zero-field two-
dimensional case. A cubic $ \pm $ $I$ system with $L = 4$ and 
periodical boundary conditions is considered in \cite{20} using 
a recursive branch-and-bound algorithm \cite{21}. All ground 
states and all low-lying excited states are characterized in 
respect to their connections in the configuration space. 
Ground-state calculations for an analogous system but using 
periodical boundary conditions in two dimensions and free 
boundaries in the third are recently given on the basis of the 
transfer-matrix method \cite{22}. It should be mentioned that 
exact ground-state calculations are complementary to Monte 
Carlo (MC) calculations (cp. \cite{23} for the three-
dimensional $\pm$ $I$ spin glass up to $L = 64$). A useful 
combination of MC methods and exact numerical investigations 
enables an analysis of the low-energy structures at least in 
local regions of the phase space, see  e.g. \cite{24}.

Spin glasses appear as suitable model systems to discuss
relaxation and aging processes in glassy systems. 
On a purely {\sl phenomenological } level the anomalously slow
dynamical behaviour is described in terms of empirical laws.
Mostly the Kohlrausch law \cite{25} is used to characterize
the time dependence of physical quantities by stretched 
exponentials. Various attempts have been made to give a 
physical understanding of this picture. The 
heuristic basis is attributed to the morphology of the phase 
space, assuming that it is partitioned in `components' by 
bifurcation-like splitting \cite{26}, which at first was 
mentioned as an idea by Krey \cite{27}. A complex spanning 
phase-space structure is suggested by the method of damage 
spreading \cite{28} analogous to a percolating cluster in a 
high-dimensional hypercube \cite{29}. Theoretical concepts on 
the basis of the infinite-range Sherrington-Kirkpatrick (SK) 
model \cite{30} support a special hierarchical topology of the 
phase space characterized by an ultrametric organization of 
metastable states which are separated by energy barriers 
\cite{31}. This picture seems to be consistent with exact 
results for systems with up to $N = 24$ spins \cite{32}, 
heuristic estimations (up to $N = 96$) \cite{33} and 
experiments \cite{34, 49}. However, there are also arguments 
against hierarchically constrained dynamics at least for 
certain time scales \cite{35}.

In {\sl mesoscopic} models special plausible assumptions on the 
phase-space structure are made. Random walks on tree structures 
\cite{36} belong to this class of theoretical concepts. Another 
picture is the droplet model (\cite{37} - \cite{39}), which seems 
to be related with the above mentioned `packet of solidary 
spins' in the $ \pm $ $I$ spin glass \cite{10}. 
Starting with the assumption of only one ground state 
(apart from the mirror symmetry of the problem)  
a droplet of the length scale $l$ is that multi-spin 
cluster of coherently flipped spins which contains a 
certain spin and belongs to the lowest excitation energy due to 
flipping \cite{38}. Such droplets were also found earlier 
in two-dimensional amorphous Ising systems with 
antiferromagnetic interactions ($N = 40 \ldots 60$) 
\cite{7, 8, 40}. Experiments 
on relaxation of the magnetization of spin glasses 
qualitatively agree with the results found on the basis of the 
droplet concept of low-lying excitations which dominate the 
long-distance and long-time correlations \cite{41}. Further 
theoretical investigations e.g. on aging in spin glasses 
\cite{42} support the droplet scenario, too. The question, 
whether or not the droplet model for short-range spin glasses 
is in contradiction with hierarchical concepts for the 
infinite-range SK model, is actively discussed \cite{49, 50}. 
Especially, the validity of the droplet theory in more than two 
dimensions is not yet clear \cite{43, 44}.

It is expected that a number of still open questions can be 
handled by means of {\sl microscopic} concepts. First 
investigations which make use of the exact knowledge of the 
low-energy landscape in the phase space of complex systems let 
this approach appear to be promising. In \cite{3} all 
configurations connected to a certain suboptimal solution of a 
Travelling Salesman problem ($N = 32$) are considered. They 
form a `phase space pocket' around this minimum. This method 
has been extended to two-dimensional ($L = 8$) and 
three- dimensional ($L = 4$) short-range Gaussian Ising 
spin glass systems \cite{45}.

The purpose of the present work is to extend the idea of 
microscopic considerations using for the first time the exact 
knowledge of the {\sl global} low-energy phase-space structure 
of a three-dimensional $ \pm $ $I$ spin glass (section 2). The 
the long-time behaviour of random walks on this structure 
represented by the largest nontrivial eigenvalue of the 
transition matrix in the case of infinite temperature is 
calculated and compared with that of an unstructured system 
(section 3).

\section{Model}

A cubic Ising spin glass system on a $ 4 \times 4 \times 4 $ 
lattice with $ \pm $ $I$ interactions between nearest 
neighbours and periodical boundary conditions in all three 
directions is considered. The distribution of interactions is 
randomly chosen with an exact portion of $50 \% $ of 
ferromagnetic and of antiferromagnetic bonds. For this system 
(1) was solved by the method of recursive branch-and-bound 
\cite{21}. Additional, all energetically low-lying states were 
calculated with this method.

In \cite{20} the mean ground-state energy per spin $ 
\overline{E}\left( T \rightarrow 0 \right) / N = 1.733 \left( 
\pm 0.013 \right) I $ and the mean ground-state entropy per 
spin $ \overline{S}\left( T \rightarrow 0 \right) / N = 0.073 
\left( \pm 0.007 \right) k_B $ ($ k_B $ - Boltzmann constant) 
were found by calculating the ground-state 
properties of 200 systems with different distributions of the 
interactions (Fig. 1).

A classification of the found states of the representative 
system in Fig. 1 by means of their neighbouring relations in 
the configuration space leads to an exact first schematic 
picture of the `valley structure' in the configuration space 
\cite{20}. This was done by arranging all the states with an 
energy lower than a chosen energy in clusters. By definition, 
two spin configurations belong to the same {\sl cluster} 
whenever a chain connecting them exists building on 
neighbouring members of this cluster. Neighbours in the 
configuration space may differ only in the orientation of a 
single spin. There are no paths between different clusters. 

At first glance the schematic picture in \cite{20} could be 
explained by a hierarchical organization of the clusters 
\cite{36, 46} as shown in Fig. 2a. 
But another more detailed investigation of the structure with 
respect to `microcanonical' clusters (i.e. all members of a 
cluster have the same energy) leads to a more complex picture 
(Fig. 2b). The found closed circle paths cannot be described by 
a hierarchy. Moreover, there are connections between different 
clusters which do not go across energy barriers but underrun 
such barriers by going first to intermediate states with lower 
energy.

It should be mentioned here that the connections drawn in Fig. 
2b only show that at least one path between the concerned 
clusters exists, but there is no information about the total 
number of paths and how difficult it is to find them. 

\section{Random Walk}

One method to quantify the connectivity of the phase space is 
the concept of the random walk. A random walk on the subset of 
$M$ low-lying states can be written in matrix notation and 
solved as an eigenvalue problem \cite{47, 48}. This will be 
outlined briefly below.

Let ${\bf p}(t)$ be the vector of probabilities, whose elements 
$p_{i}(t) $ give the probability, that at time $t$ the system 
is in the state $i$. One time step in the random walk will then 
be described by
\begin{equation}
{\bf p}( t + 1 ) = {\bf A p}(t)
\end{equation}
where {\bf A} is the transition matrix with
\begin{equation}
A_{ij} = \left\{ 
\begin{array}{cl}
w_{ij} & {\rm for} \; i \neq j \\
\parbox[]{3cm}{\[ 1 - \sum_{k = 1, k \neq i}^M w_{ki} \] }& 
{\rm for}\; i = j
\end{array} \right. \, .
\end{equation}
The element $ w_{ij} $ denotes the transition probability per 
unit time from state $j$ to state $i$ and is given by
\begin{equation}
w_{ij} = \left\{
\begin{array}{cl}
0 & $ {\rm if states i and j are not nearest neighbours}$ \\
 & ${\rm (i.e. Hamming distance $ > 1$)}$ \\
\frac{1}{N} & $ {\rm if states i and j are nearest neighbours 
and} $ E_i \leq E_j \\
\frac{1}{N} \exp\{- \beta (E_i - E_j)\} & $ {\rm if states i 
and j are nearest neighbours and} $ E_i > E_j
\end{array} \right. \, ,
\end{equation}
where $ \beta = 1 / k_B T $, $ E_i $ denotes the energy of 
state $i$ and $N$ is the maximal number of nearest neighbours, 
i.e. the number of spins. Because of the definition of {\bf A} 
the total probability in the system is constant.

Only for $ \beta = 0 $ is the matrix {\bf A} symmetric. For $ 
\beta > 0 $ the problem can be transformed into a symmetric one 
taking into consideration the detailed balancing in the 
equilibrium distribution \cite{47}. So in the following we 
restrict ourself to the case $ \beta = 0 $.

For calculating the time evolution of the system it is useful 
to split any starting state at time $ t = 0$ by the system of 
eigenvectors of {\bf A}:
\begin{equation}
{\bf p}(0) = \sum_{i=0}^{M-1} \alpha_i {\bf b}_i \; , \; \; \; 
{\bf A b}_i = \lambda_i {\bf b}_i \; \; \; (i = 0 \ldots M-1) 
\, .
\end{equation}
Because the matrix {\bf A} is real and symmetric, all $ 
\lambda_{i} $ must be real and can be ordered by $ \lambda_0 
\geq \lambda_1 \geq \ldots \geq \lambda_{M-1} $. Furthermore 
all eigenvalues are restricted to $ | \lambda_i | \leq 1 $. The 
largest eigenvalue $ \lambda_0 = 1 $ is the trivial one and 
describes the equilibrium state $ {\bf b}_0 = {\bf p}(\infty)$ 
with $ p_i(\infty) \propto \exp(- \beta E_i) $ .

Rewriting (2) with respect to (5) the probability distribution 
at time $t$ is given by
\begin{eqnarray}\nonumber
{\bf  p}(t) & = & {\bf A p}(t-1) = {\bf AA} \ldots {\bf Ap}(0) 
= {\bf A}^t {\bf p}(0) = {\bf A}^t \sum_{i = 0}^{M-1}\alpha_i 
{\bf b}_i \\
& = & \sum_{i = 0}^{M-1} \lambda_i^t \alpha_i {\bf b}_i \, .
\end{eqnarray}

A simple measure of the spreading in the configuration space is 
the mean distance $ r_{k}(t)$ from the starting configuration 
$k$ of a random walk. If $ {\bf h}_{k} $ denotes the vector of 
Hamming distances to the configuration $k$, $ r_{k}(t) $ can be 
written as $ r_k (t) = {\bf h}_k^T {\bf p}(t) $ . The deviation 
$ q_{k}(t) $ to the equilibrium distance is then
\begin{eqnarray} \nonumber
q_k(t) & = & r_k(\infty) - r_k(t) = {\bf h}_k^T \left( {\bf 
b}_0 - \sum_{i = 0}^{M-1} \alpha_i {\bf b}_i \lambda_i^t 
\right)\\ & = & - \sum_{i=1}^{M-1} \alpha_i {\bf h}_k^T {\bf  
b}_i \lambda_i^t = \sum_{i=1}^{M-1} \gamma_{ki} \exp(t \, {\rm 
ln} \lambda_i) \; ; \; \; \gamma_{ki} = \alpha_i {\bf h}_k^T 
{\bf h}_i
\end{eqnarray} 
using  $  \alpha_{0} = 1 $. In the limit of large times one 
gets
\begin{equation}
q_k(t \rightarrow \infty) \propto \exp (t \, {\rm ln} 
\lambda_1) \, .
\end{equation}

Fig. 3 shows $ q_{k}(t) $ for random walks in the 
representative system with different starting configurations 
$k$ and $ \beta = 0 $. The considered subset consists of all 
states with an energy lower or equal to the second excitation. 
As expected for long times all curves show an exponential decay 
with the same exponent ln$\lambda_1 = -2.98 \cdot 10^{-6} $, 
which does not depend on the starting position. The time 
belonging to this value $ \tau_1 = 3.36 \cdot 10^5 $ is 
considerable larger than that of an unstructured system, which 
was calculated in a comparing calculation as $ \tau_1^{\ast} = 
74 $, see Appendix A.

\section{Conclusions}

Based on the exact knowledge of the global 
low-energy phase-space structure for a 
$ 4 \times  4  \times  4 \; \pm $ $I$ 
Ising spin glass it is possible to obtain a schematic detailed 
picture, which shows clusters in the phase space and their 
connections. The appearence of closed circle paths and the 
possible underrunning of energy barriers are difficult to 
describe solely by a simple hierarchical concept. 

By the reason of this complicated structure, in general, the 
relaxation behaviour should be determined not only by the 
existing energy barriers but also by `entropic barriers' 
\cite{26}. In the case of increasing temperature the influence 
of the energetic barriers should decrease. Thus, if there is 
any entropic effect it should be found first in the limes of 
infinite temperature. However, it should be noted that there is 
an influence of the cut-off energy, which is necessary in our 
calculations.

In first approximation the long-time behaviour of a random walk 
in the low-energy landscape of the system can be characterized 
by the largest nontrivial eigenvalue of the transition matrix. 
It could be shown that in comparison with an unstructured 
system there is a slowing down by a factor of about $ 10^{3} $ 
in the relaxation. 

The additional temperature-dependent influence of  energy 
barriers on the dynamics is the subject of  further 
investigations.

\section*{Acknowledgement}

We would like to thank I. A. Campbell for fruitful discussion 
and constructive comments. This work was supported by the DFG 
(project no. Ko 1416).

\begin{appendix}
\section*{Appendix A - Unstructured Systems}
An unstructured system in our sense is a system with minimal 
structural information. For the reason of comparing it with our 
system it should have the same number of states $M$. For every 
state of the unstructured system the number of nearest 
neighbours $d$ has to be equal to the mean number of nearest 
neighbours in our system. The number of spins $N$ and thus the 
dimension of the hypercube describing the whole configuration 
space is equal for both systems. 

The construction of a fictive approximated unstructured system 
was done by dilution of a $D$-dimensional hypercube where $ D  
\leq N $. This hypercube is embedded in the large hypercube 
with dimension $N$. Every state of the subcube has to be 
occupied with a probability of  $d/D$ to get $d$ nearest 
neighbours instead of $D$ for the fully occupied subcube. Since 
for $d \geq 1$ this probability is larger than the percolation 
threshold of such a hypercube \cite{29} the remaining subset 
should form at least one large connected cluster.

Starting the random walk from state $i$ because of the symmetry 
of the system the occupation probability of state $j$ should 
only depend on the distance between these two states. Therefore 
the states of the diluted subcube can divided into $D+1$ layers 
where the occupation probability of a state in the $k$-th layer 
with Hamming distance $k$ to the starting configuration at time 
$t$ is denoted by $ p_k(t) $. 

The number of states in the $k$-th layer can be calculated as 
\begin{equation}
n_k = \frac{d}{D}\left( \begin{array}{c} D \\ k \end{array} 
\right) \, .
\end{equation}

If $c_k^+ $ and $ c_k^- $ are the numbers of nearest neighbours 
for a state of the $k$-th layer in the $(k+1)$-th layer and the 
$(k-1)$-th one, respectively, it follows that
\begin{equation}
c_k^+ + c_k^-	=d \, .
\end{equation}

The number of all connections between layer $k$ and $k+1$ must 
be the same as between layer $k+1$ and $k$ and therefore
\begin{equation}
n_k c_k^+ = n_{k+1} c_{k+1}^- = n_{k+1} \left( d-c_{k+1}^+ 
\right) \; \Rightarrow \; c_{k+1}^+ = d - \frac{k+1}{D-k} c_k^+
\end{equation}
and
\begin{equation}
c_0^+ = d \Rightarrow c_1^+ = d \frac{D-1}{D} \Rightarrow 
\ldots \Rightarrow \underline{c_k^+ = d \frac{D-k}{D}} \; 
\Rightarrow \; \underline{c_k^- = d \frac{d}{D}} \, .
\end{equation}
Now the occupation probability of a state in the $k$-th layer 
at time $t+1$ can be written as
\begin{eqnarray} \nonumber
p_k(t+1) & = & p_k(t) \left( 1 - \frac{d}{N} \right) + p_{k-
1}(t) \frac{c_k^-}{N} + p_{k+1}(t) \frac{c_k^+}{N} \\
 & = & p_k(t) \left( 1 - \frac{d}{N} \right) + p_{k-1}(t) 
\frac{d k}{D N} + p_{k+1}(t) \frac{d (D-k)}{D N} \, .
\end{eqnarray}

Eq. (13) is equivalent to a matrix multiplication $ {\bf 
p}(t+1) = {\bf B p}(t) $ where the matrix {\bf B} is given by
\begin{equation}
B_{kj} = \left\{ \begin{array}{cl}
\frac{d (D-k)}{DN} & ${\rm if } $ j = k - 1 \\
1- \frac{d }{N} & ${\rm if } $ j = k \\
\frac{d k}{DN} & ${\rm if } $ j = k + 1 \\
0 & ${\rm otherwise}$
\end{array} \right.
\, .
\end{equation}

The long-time behaviour of the unstructured system is given by 
the largest nontrivial eigenvalue of {\bf B}. For a given set 
of $N$, $D$ and $d$ this can easily be done using numerical 
standard methods.

The values of $N$ and $d$ are defined by the number of spins 
and the mean number of nearest neighbours in the configuration 
space of our representative system ($N = 64$, $d \approx  
7.7$). $D$ has to be chosen in such a way that the total number 
of states is correct (i.e. $\approx 10^{5}$ in our case). The 
total number of states in the unstructured system is
\begin{equation}
M = \frac{d}{D} \sum_{k=0}^D \left( \begin{array}{c} d \\ D 
\end{array} \right) = \frac{d}{D}\, 2^D \; \Rightarrow \; 
\underline{D \approx 18} \, .
\end{equation}

Using this combination of $N$, $D$ and $d$ the eigenvalue can 
be calculated as $ \lambda_{1} = 0.987 $ and ln$\lambda_{1} = -
1.4 \cdot 10^{-2}$. The time belonging to this values is $ 
\tau_1^{\ast} = 74 $.
\end{appendix}

\newpage

\large{ \bf Figures}
\begin{enumerate}
\item{\label{Fig 1.}Energy and degeneration of ground states 
for 200 random systems.}
\item{\label{Fig 2.}Detailed structure of the configuration 
space up to and including the second excitation for the 
representative system. The drawn circles denote microcanoncial 
clusters. For each excitation the size of a circle is 
proportional to the number of states in the cluster, where the 
scaling factor differs for different excitations.\\
a) Explanation by a schematic picture of a hierarchical 
structure \\
b) Result of an exact detailed analysis (the full lines mark an 
example of a closed circle path).}
\item{\label{Fig 3.} $q_{k}(t)$ for random walks in the 
representative system with different starting configurations 
$k$ for $ \beta = 0 $.}
\end{enumerate}

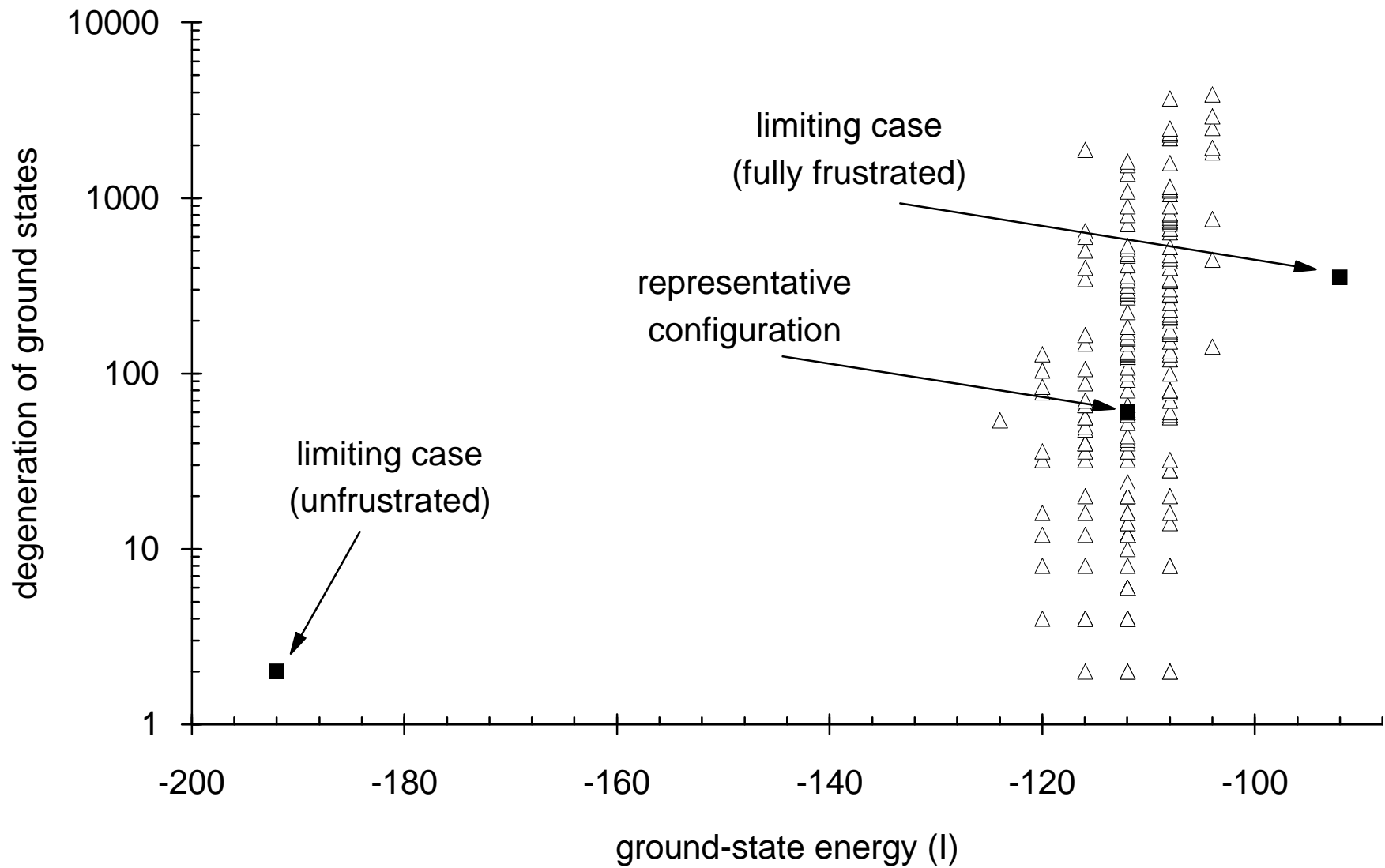
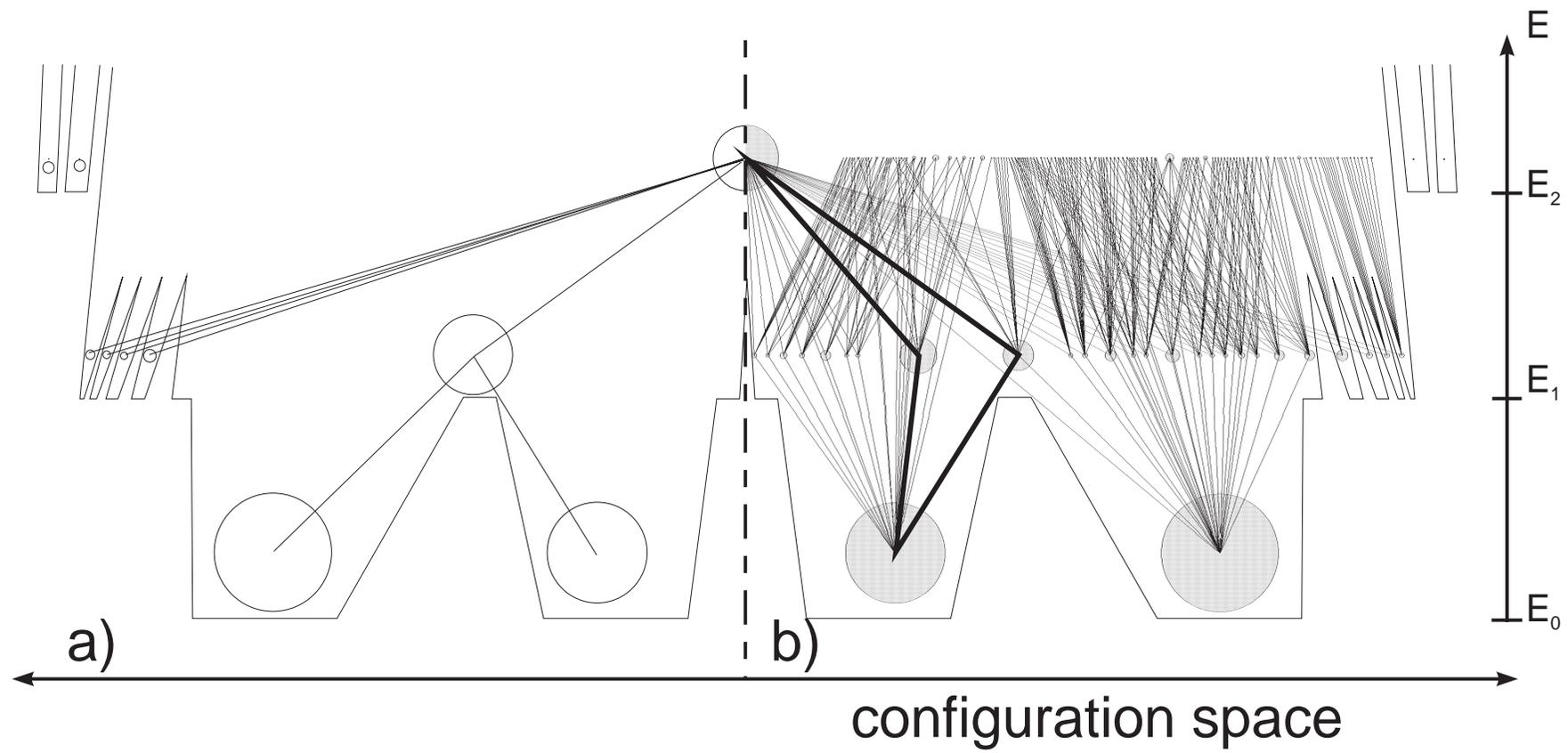
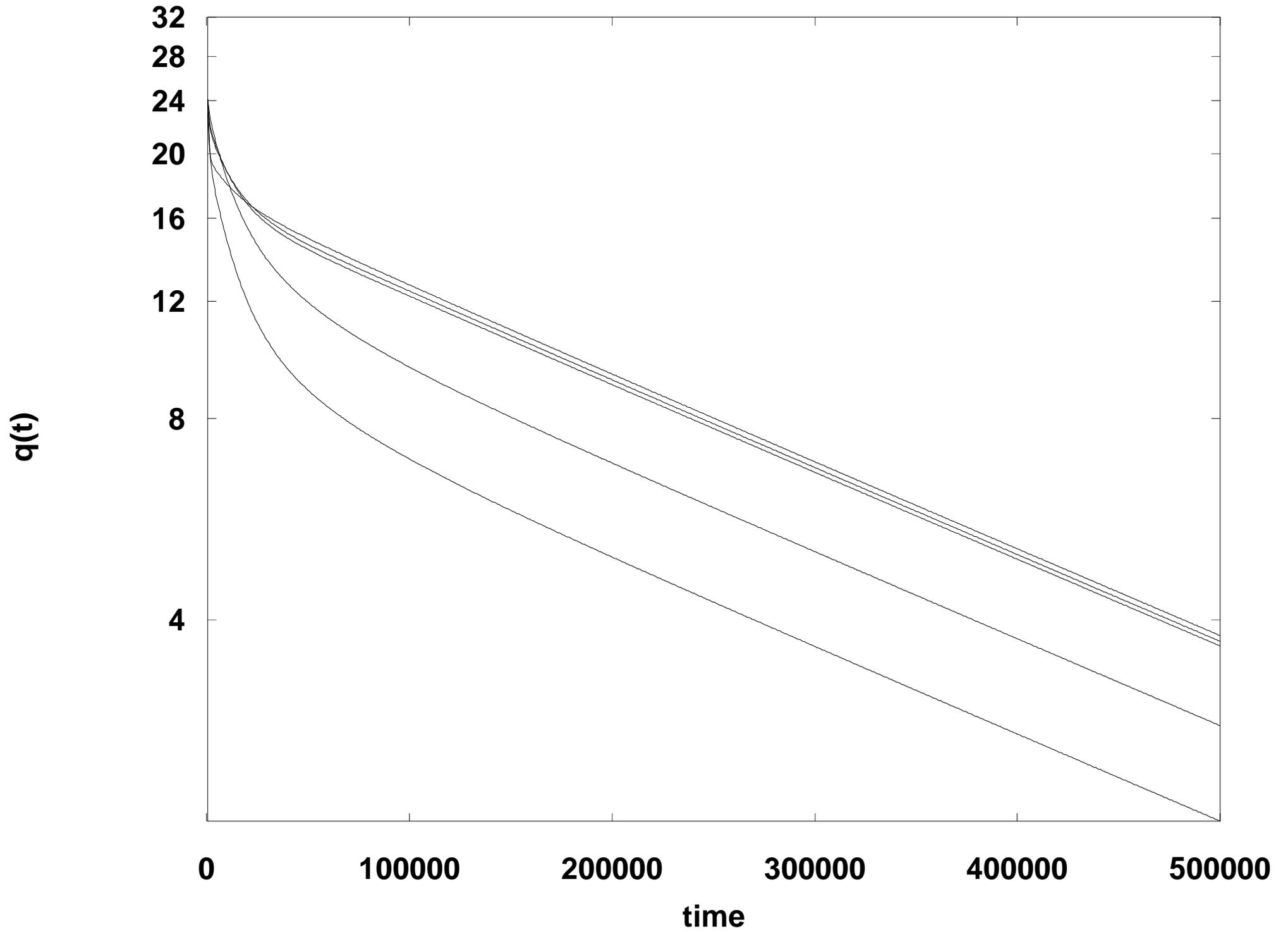
\end{document}